\documentclass[12pt]{article}

\begin {document}
$$ $$
\vfill

\begin{center}
{{\bf\LARGE
Relativistic Quantum Non-Locality}}

\bigskip
Erast B. Gliner\footnote{Visiting physicist at Stanford Linear
Accelerator Center, Stanford University, California 94309} \\
I.Th.S., 661 Faxon Avenue, San Francisco, CA 94112\footnote{
E-mail:  erast@hotmail.com}

\end{center}

\vfill

\begin{center}
{\bf\large
Abstract }
\end{center}

The controversy between relativistic causality and quantum non-locality
can be resolved by establishing the general relativistic background of
quantum non-locality.

\vfill

\begin{center}
\end{center}

\vfill

\vfill\eject


\normalsize

\section{Introduction}

The relaxation of restrictions, imposed by the relativistic causality on
links between distant physical events, seems to have no reasonable
alternative in many fields of physics.  In cosmology, the synchronous
start of expansion is hard to explain without an instantaneous
omnipresent initiation.  Such a problem persists, {\em e.g.}, in the
model based on Sakharov's idea of a non-singular initial
state~\cite{ref:1}, as well as in inflationary models, which suggest the
synchronous start of the phase transition at the end of inflation.  On
the other end of scale, the relativistic causality is challenged by the
quantum non-locality.

This challenge is the subject of the discussion below aimed to merge
quantum non-locality with relativistic physics.

\section{Time arrow representation}

The spacetime {\em history of a quantum object} is built up by two kinds
of events:  {\em interactions} at the {\em intersections} of histories,
and {\em free falling}, with no disturbance reaching the object.

Quantum mechanics describes a free falling object by a set of {\em
constants}, quantum numbers and mechanical integrals of motion, defined
by quantum and relativistic {\em conservation laws}.  At intersections,
the conservation laws generally leave room for the redistribution of
conserved variables among the products of interactions.  The final
distribution is predicted, at best, only stochastically.

The source of stochasticity can be formally described as the exposure of
quantum objects to {\em random fluctuations} obeying the uncertainty
principle.  Conservation laws strictly suppress the effect of
fluctuations on a free falling object:  in the average, all the
uncertainties must vanish (just this lets us see remote celestial
objects).  But at the intersections the fluctuations can randomly affect
interactions.  This distinguishes a free falling quantum object from an
interacting one.

In principle, the randomness attributed to fluctuations could be driven
by some deterministic machinery.  (Recall, {\em e.g.}, the infinite
decimal expression for the number $\pi$:  the sequence of figures is
perfectly random, though each figure is strongly determined.)  Far from
being resolved, the issue of determinism is still out of reach of
contemporary theory, and the real concern is reversibility rather than
causality.

Because of the time-symmetry of quantum formalism, we admit that it must
be applied to a direct physical process and to its reversal identically.
This means that if the theory defines a physical process stochastically,
the same is true for its reversal, with the reversed process being
dependent only on its {\em past} in terms of the reversed time ({\em
cf.} Ref.  \cite{ref:2}).  We will call a process {\em reversible} if
the reversal of the time direction ({\em time reversal} for short),
being employed {\em twice}, returns exactly the original process.  It
follows then from the quantum formalism that, contrary to the explicit
reversibility of free falling histories, the {\em intersections} of the
space-time histories are {\em generally irreversible} because of the
stochasticity of quantum formalism.  This means that {\em some aspects
of preexisting reality are theoretically unrecoverable}, and the {\em
subsequent evolution is not completely predictable}.

Since the information lost during interactions is not restored by time
reversal, we may admit that it is the loss of information that {\em
inside intersections} creates the {\em local distinction between the
past and the future}, making cause and effect non-interchangeable
physically.  Such occurrences can be labeled as the {\em arrow of time}.

The intersections, however, do not cover the whole spacetime.  What then
in the remaining part of spacetime---let us call it {\em open}
spacetime---creates the local distinction between past and future, {\em
i.e.}, between cause and effect?  The striking answer is ``{\em
nothing}''.  The {\em constants}, determining the state of a
free-falling object, do not specify the arrow of time.  The spacetime
geometry is alien to the notion of direction, so that no geometrical
means exists for delivering the time arrow from afar.  The fact that
free falling histories are future-directed in the global reference frame
is not relevant, because the principle of relativity denies the
influence of relative velocity on local physics.

Thus, in open spacetime a quantum object is free of the arrow of time.
We face the alternative:  either this is a kind of easily removable
degeneration, which is unnoticeably eliminated in any interaction, or
this is the inherent property of free falling quantum objects, and then
the way in which such objects behave is radically distinct from that of
free falling macroscopic bodies.

Decoherence, prevailing in the microscopic structure of the latter,
involves a lot of {\em causally} related properties, such as
irreversibility and the arrow of time.  Based on these properties, the
propagation of a disturbance through a macroscopic body proceeds along
the time arrow with the speed of sound.  In the absence of the time
arrow, the notion of propagation, as a continuous sequence of the cause
and effect, becomes self-contradictory.  Thus, physics without the arrow
of time is also deprived of the cause and effect propagation.  So,
entering interactions, a free falling quantum object can behave only as
an {\em indivisible whole}.  The consideration of the {\em distant
quantum objects} in the next section reveals just this type of behavior.

\section{Deon, distant entangled object}

A quantum object is called {\em entangled} if its wave function is not
the product of the wave functions of its components.  If its size is
much larger than the sizes of its components, it is called a {\em
distant entangled object}; we will call it {\em deon} for short.

Consider Bohm's version~\cite{ref:3} of the well-known
EPR-thought-experiment~\cite{ref:4}, which historically turned out to be
the first challenge to relativistic causality.

An emitter in each of its working cycles shoots out in opposite
directions along its axis a pair of spin-1/2-particles, with total spin
zero, but with the spin of each particle remaining undetermined.  The
last condition implies that the particles {\em share} the zero spin of
the pair, and this creates some kind of interdependence between them.
Only a pair as a whole is an independent quantum object, which is a deon
with the growing distance between its constituents.

The EPR-deon is one of the innumerable quantum entangled objects, whose
constituents share some quantum numbers and display the behavior known
as {\em quantum non-locality}.

Let, on each side of the emitter, be stationed an observer who measures
a spin component of the approaching particle.  For measurements, each
observer independently chooses one of the two predetermined mutually
orthogonal directions, normal to the emitter axis.  Quantum formalism
predicts that the stochastic distribution of readings, found by {\em any
single} observer, depends neither on the independent choice made by the
other, nor on the fact that the particles belong to a deon.  This means
that the {\em particles' link with a deon does not affect local
physics}.  In particular, EPR-deon cannot be a means of communication
between observers.

The occurrence of deons can be revealed only {\em post factum} by
analysis of the correlation between readings pertaining to the {\em
pairs} of particles.  The cases, when the measurements, related to the
same pair, have been made in {\em distinct} directions, reveal no
correlation, but the same choice of direction {\em always yields the
opposite results}, and this discloses the presence of the deon, {\em
i.e}, a quantum number shared by both particles.

In terms of quantum formalism, this pattern can be explained only if 
the measurement of  the spin of one particle somehow changes 
the wave function of the twin particle, which also acquires the
definite spin component.  The latter is  opposite in direction to that found 
for the first particle.  Then and only then the required pattern appears, and angular
momentum is conserved.

The mechanism of this process is, however, obscure.  To keep the angular
momentum unchanged, the interaction between the deon constituents either
should be instant, and then the process cannot be described in terms of
relativistic causality, or quantum formalism is incomplete, and then it
can be expanded to include a carrier, which transports action from one
deon constituents to another ({\em cf.} Ref.~\cite{ref:4}).

The seven-decade attempts to incorporate the last idea into quantum
mechanics, however, failed.  Thus, {\em quantum non-locality}---quantum
{\em instant} (or, in other words, {\em spacelike}) propagation---is,
most likely, in the nature of things.  Though this is seemingly in
variance with relativistic causality, the discrepancy is {\em
conceptual} rather than {\em physical}, because, just as for the EPR
deon, the {\em local physics is never affected}, and by means of that
{\em no actual violation of relativistic causality takes place}.

The simplest suggestion for resolving the problem is that the place,
where relativistic causality is in force, is separated from that where
the quantum non-locality can be observed.  The suitable places are open
spacetime and intersections of histories.

The expected deon behavior supports this idea.  At the intersections of
histories with massive redistribution of conserved variables, multiple
deons can appear; each of them being a cluster of free falling
particles, which share some quantum numbers.  When a multiple deon runs
into an intersection, the particles, which are actually involved in
interactions, instantly acquire definite quantum numbers, and so do
their twins.  Thus, due to these instant adjustments, the deon as a
whole escapes interactions.  Only former constituents that have broken
off with the deon are actually involved in the interaction.
Constituents, keeping their non-local quantum ties intact, remain in open
spacetime where they form the altered deon.  
Deons therefore exist merely in open spacetime as free
falling objects, and therefore escape from the incompatibility with the
relativistic causality acting merely inside intersections of histories.
It is worth emphasizing that this separation of powers is essentially
based on the quantum non-locality.

\section{Relativistic quantum non-locality}

A causal interaction is continuously decomposable into {\em local}
cause-and-effect relations along paths confined to a light cone.  Since
quantum non-locality does not affect local physics, it cannot be
displayed in this way.  This, in classical terms, looks like the {\em
action at a distance}.  The {\em Lorentz invariant} counterpart of the
latter is {\em tachyon mechanics}, the internally consistent
superluminal paraphrase of special relativity~\cite{ref:5}.  This
implies that quantum non-locality should obey tachyon mechanics and
therefore is Lorentz invariant.

As was shown long ago~\cite{ref:6}, tachyonlike faster-than-light links
between physical events can be incorporated into quantum formalism
without causality breaking, but only if tachyons are virtual and never
appear as free propagating particles.  This essential restriction is actually
inherent in tachyon mechanics itself.  Indeed, let $u$ be the velocity
of a particle and $\Delta t$ the time interval between its emission and
absorption measured by an inertial observer.  Lorentz transformations
indicate that another inertial observer, moving in the same
space-direction with the relative velocity $v$, finds this time interval
to be $(c = 1)$:
\[
\Delta t^\prime = \frac{1-uv}{\sqrt{1-v^2}}\, \Delta t \ .
\]
If the particle is a tachyon, {\em i.e.} $ u>1$, and the velocity
$v> u^{-1}$, then
$\Delta t$ and $\Delta t^\prime$ are of opposite signs.  This means,
that what one observer sees as emissions, another does as absorption,
and vice versa.  This is in striking contrast to relativistic causality,
where it is of the prime importance that the sequence of cause and
effect, or in other words the arrow of time, is Lorentz invariant.  This
means that an invariant time arrow cannot be introduced in tachyon
mechanics, {\em i.e.} the latter does not describe the real propagation
of anything.  (The tachyon propagation is spacelike, so that in terms of
causality it deals with the already existing relations between things.)
This feature of tachyon mechanics is, however, completely in line with
above-considered properties of non-locality.

Thus, we have at hand:  The concept of open spacetime deprived of the
arrow of time and by means of that providing the room for non-locality.
The Lorentz invariant tachyon mechanics that describes the spacetime
properties of non-locality.  Quantum non-locality---the collection of
supporting facts that fills up the {\em still empty} tachyon niche in
relativistic physics.  These three ingredients seem to be the foundation
of {\em the relativistic quantum non-locality}.  Causality survives, and
non-locality comes as a part of the same relativistic physics, which is
responsible for causality.  This brings to a close the debate on the
conflict between quantum non-locality and relativistic causality.

At present, we can only guess the possible physical role of the
relativistic quantum non-locality.  We cannot rule out that it is the
spacelike links, introduced by the non-locality, which are responsible
for the simultaneous start of the cosmological expansion, as well as for
the homogeneity and basic hallmarks of the universe arising from the
initial state with all conserved quantities globally shared.

\section*{Acknowledgments}

The author thanks {\em The Andrei Sakharov Foundation} for the
literature grant, {\em Stanford Linear Acceleration Center} for
hospitality, and Michael E. Peskin for stimulating criticism.  The
author gratefully acknowledges the recent works by Stephen Hawking,
Roland Omnès, Roger Penrose, Huw Price, L. S.\  Shulman and many others,
which with modern inquisitiveness make it clear that no deciphering of
the quantum reality allows avoiding the exploration of the spacelike
paths in general relativity.  As a matter of fact, this final victory of
quantum thinking is coming without sacrificing Einsteinian relativity.


\begin{thebibliography}{99}

\bibitem{ref:1}
A. D.  Sakharov, Zh.  Eksp.  Teor.  Fiz. {\bf 49}, 345  (1965);
E. Gliner and 	I.  Dymnikova, Sov.  Astron.  Lett. {\bf 1}, 93  (1975).

\bibitem{ref:2}
Huw Price (1995), in {\em Time's Arrow Today}, Steven F. Savitt ed.,
Cambridge University Press.

\bibitem{ref:3}
D.  Bohm, {\em Quantum Theory}, New York:  Prentice Hall  (1951).

\bibitem{ref:4}
A.  Einstein, B. Podolsky and N. Rosen, Phys.  Rev. {\bf 47}, 777  (1935).

\bibitem{ref:5}
E.  Wigner, Ann.  Math. {\bf 40}, 149 (1939).

\bibitem{ref:6}
S.  Tanaka, Progr.  Theor.  Phys. {\bf 24}, 171 (1960); G. Feinberg, Phys.
Rev. {\bf 159}, 1089 (1967).

\end{thebibliography}
\end{document}